\documentclass[aps,pra,twocolumn,groupedaddress,nofootinbib,notitlepage,showpacs,floatfix,superscriptaddress]{revtex4-1}
\usepackage{graphicx}% Include figure files
\usepackage{dcolumn}% Align table columns on decimal point
\usepackage{amsthm}
\usepackage{bm}% bold math

\pdfoutput=1
\usepackage{graphicx,graphics,epsfig,subfigure,times,bm,bbm,amssymb,amsmath,amsthm,mathrsfs,MnSymbol}
\usepackage{gensymb}
\usepackage{amsfonts}
\usepackage[matrix,frame,arrow]{xypic}
\usepackage[pdfstartview=FitH]{hyperref}
\hypersetup{
    colorlinks=true,       % false: boxed links; true: colored links
    linkcolor=red,          % color of internal links
    citecolor=magenta,        % color of links to bibliography
    filecolor=magenta,      % color of file links
    urlcolor=cyan,           % color of external links
    runcolor=cyan
}
\usepackage{epstopdf}
\usepackage[pdftex]{color}
\usepackage{hypernat}
\usepackage{braket}%Dirac Notation in QM
\usepackage{enumerate}
\usepackage[normalem]{ulem}
\usepackage[usenames,dvipsnames]{xcolor}
\usepackage{multirow}
\usepackage{mathtools}

\definecolor{orange}{rgb}{1,0.5,0}

\newcommand{\ignore}[1]{}
\usepackage{geometry}

\newcommand{\ud}{\mathrm{d}}

\ignore{
\documentclass[eprintnumbers,amsmath,amssymb,onecolumn,a4paper,caption ]{article}
\usepackage{amsfonts}
\usepackage{amssymb}
\usepackage{mathrsfs}
\usepackage{mathbbold}
\usepackage{bbm}
\usepackage{mathrsfs}
\usepackage{dcolumn}% Align table columns on decimal point
\usepackage{bm}% bold math
\usepackage{times,epsfig,amssymb,amsmath}
\usepackage{float}
\usepackage{subfigure}
\usepackage{geometry}\geometry{left=2.5cm,right=2.5cm,top=3cm,bottom=3cm}

\usepackage{color}

}

\begin{document}

\title{Diagonal Entropy in Many-Body Systems: Volume Effect and Quantum Phase Transitions}% Force line breaks with \\

\author{Zhengan Wang}
\affiliation{Institute of Physics, Chinese Academy of Sciences, Beijing 100190, China}
\affiliation{School of Physical Sciences, University of Chinese Academy of Sciences, Beijing 100190, China}

\author{Zheng-Hang Sun}
\affiliation{Institute of Physics, Chinese Academy of Sciences, Beijing 100190, China}
\affiliation{School of Physical Sciences, University of Chinese Academy of Sciences, Beijing 100190, China}

\author{Yu Zeng}
\affiliation{Institute of Physics, Chinese Academy of Sciences, Beijing 100190, China}
\affiliation{School of Physical Sciences, University of Chinese Academy of Sciences, Beijing 100190, China}

\author{Haifeng Lang}
\affiliation{Institute of Physics, Chinese Academy of Sciences, Beijing 100190, China}
\affiliation{School of Physical Sciences, University of Chinese Academy of Sciences, Beijing 100190, China}

\author{Qiantan Hong}
\affiliation{School of Physics, Peking University, Beijing 100871, China}

\author{Jian Cui}
\email{jianCui@buaa.edu.cn}
\affiliation{School of Physics, Key Laboratory of Micro-Nano Measurement-Manipulation and Physics (Ministry of Education), Beihang University, Beijing 100191, China}

\author{Heng Fan}
\email{hfan@iphy.ac.cn}
\affiliation{Institute of Physics, Chinese Academy of Sciences, Beijing 100190, China}
\affiliation{School of Physical Sciences, University of Chinese Academy of Sciences, Beijing 100190, China}
\affiliation{CAS Central of Excellence in Topological Quantum Computation and Collaborative Innovation Center of Quantum Matter, Beijing 100190, China}

\begin{abstract}
We investigate the diagonal entropy(DE) of the ground state for quantum many-body systems, including the XY model and the Ising model with next nearest neighbour interactions. We focus on the DE of a subsystem of $L$ continuous spins. We show that the DE in many-body systems, regardless of  integrability, can be represented as a volume term plus a logarithmic correction and a constant offset. Quantum phase transition points can be explicitly identified by the three coefficients thereof. Besides, by combining entanglement entropy and the relative entropy of quantum coherence, as two celebrated representatives of quantumness, we simply obtain the DE, which naturally has the potential to reveal the information of quantumness. More importantly, the DE is concerning only the diagonal form of the ground state reduced density matrix, making it feasible to measure in real experiments, and therefore it has immediate  applications in demonstrating quantum supremacy on state-of-the-art quantum simulators.
\end{abstract}

\maketitle

\clearpage

\section{Introduction}

The development of quantum mechanics is leading to emergent quantum technologies, where ``quantumness" in terms of coherence and entanglement plays a crucial role \cite{Ladd2010}.
As a prominent manifestation of quantumness, the physical concept of quantum coherence essentially roots in the superposition principle of quantum mechanics to characterize the wave-like nature of physical objects, i.e., the state of matter can be described as the superposition of at least two split states of the same object. In this sense, quantum coherence features the quantumness of single party, regardless of single body or many-body system. For the latter the single party can be either the many-body system as a whole or part of it.
While the notion of quantum coherence has long been utilized in quantum optics \cite{Glauber1963,Sudarshan1963,Kim2002}, its mathematical framework has not been developed until recently in Refs. \cite{Levi-Mintert2014,Plenio2014,Girolami2014}, where resource theories were rigorously established.
In this way, suitable quantifications of quantum coherence have been conceived including those based on relative entropy, the $l_1$ norm and many others \cite{Winter2016,MaXiongFeng2015,Streltsov2015,Piani2016}.
Entanglement also stems from superposition, however, it features inter-party relationship/correlation which is beyond the realm of classical world. As such, each superposition state has to represent bipartite or multipartite system by definition. Opposite to coherence, quantifications of entanglement were established much earlier and investigated quite widely \cite{EntanglementVedral,EntanglementVidal,EntanglementHorodecki}.

For many-body systems both coherence and entanglement are physically valid. It is then important to compare the two representatives of quantumness \cite{Streltsov2015,JinXianMin2017}. 
We focus on the bipartite scenario, i.e., the whole system with $N$ qubits is divided into two parties: subsystem $A$ of $L$ qubits and subsystem B with the other $N-L$ qubits. For pure state, entanglement between $A$ and $B$ is measured by the von Neumann entropy $S(\rho_{A})$, where $\rho_A$ is the reduced density matrix for $A$. Regarding coherence, we consider the quantification based on relative entropy, and thus the coherence of party $A$ is quantified as $C(\rho_A)=S(\rho_A^{\rm diag})-S(\rho_A)$ \cite{Plenio2014}, where $\rho_A^{\rm diag}$ is the diagonal part of $\rho_A$ obtained by removing all its off-diagonal entries.
The ``diagonal entropy" (DE) is obtained as the summation of quantum entanglement and quantum coherence
\begin{eqnarray}
S(\rho_A^{\rm diag}) = S(\rho_{A}) + C(\rho_A).
\label{Eq:DE}
\end{eqnarray}
Notice that $S(\rho_A^{\rm diag})$ is basis dependent.
The DE was first introduced in Ref. \cite{firstDE} and then carefully studied in Refs. \cite{IKEDA2015338, PREDE} by choosing the energy eigenbasis to explore thermodynamics properties of isolated quantum systems out of equilibrium. In this paper, however, we focus on the quantum feature of many-body states/systems, and hence the energy eigenbasis is not the best option. In what follows we mainly choose the eigenstates of $\sigma^z$ operators as the basis for spin-$1/2$ systems to adapt to most experimental measurements. Note that the actual values of DE may vary for other bases, nevertheless the conclusions are unaffected. For instance we will show through the XY model that although for each given ground state in the phase diagram the diagonal entropy measured with different bases (i.e., $\sigma_x$ basis and $\sigma_z$ basis) differs, the non-analytic point remains at the same location.
Note that since DE is the sum of two non-negative quantifications of quantumness, it should reflect certain quantum characteristics of the system.

In this paper we establish the connection between the scaling formula of DE and quantum phase transitions. In particular we argue that in general the leading term of the DE for a many-body ground state should obey volume law. We show through two species of examples that the predicted DE volume effect as well as the logarithmic correction is strictly followed, irrespective of the (non)integrability of the systems, and the corresponding quantum phase transition points can be accurately located by the fitted coefficients. 

\section{Formulation of DE and quantum phase transitions}\label{S:DEandQPT}
A quantum phase transition originating from quantum fluctuations may occur at zero temperature. Over the past years, many types of quantum
properties of ground states (GSs), in particular quantum entanglement and quantum coherence among other non-classical correlations,
have been successful in distinguishing different quantum phases,
see, for example, Refs. \cite{VedralRMP,AmicoNature,Nielsen,Vidal,Korepin,FanPRL,Latorre,Cardy,Wilczek,Klich,Refael,JianCui,Alcaraz,Moore,exp3,ChenJJ}.
It is believed that different quantum phenomena including quantum phase transitions might be identified by the characteristics of the entanglement or other non-classical quantum correlations of the GSs and fields.  Entanglement entropy (EE) thus has been extensively studied for many-body systems, especially when area law applies \cite{arealaw,ZanardiArea,PlenioArea,Hastings,Cirac}.
Additionally, the profoundly entangled GS can also embrace topological
EE on occasions, which characterizes the global feature of the topological order for the system where the sub-leading term of EE for topologically ordered state
can lead to topological EE \cite{Kitaev,WenXG}.
Our goal is to find the connection between quantum phase transitions and the quantum characteristic behavior of DE in analogy to yet beyond that of EE, thus establishing DE as a powerful tool to investigate various quantum phenomena.

Ground-state EE for a block of contiguous $L$ spins of one-dimensional spin chains with short-range interactions demonstrates two different behaviors depending on the energy gap \cite{Vidal,FanPRL,VedralRMP}.
For a gapped model, EE approaches a constant, i.e., EE area law is strictly obeyed, while for a gapless model, EE grows logarithmically with $L$, and the prefactor of the logarithm term is a constant related with the central charges of the corresponding conformal field theory of the model \cite{Vidal}.
As a combination of EE and coherence, DE should generally contain all terms in EE.
On top of that, quantum coherence contributes to the diagonal entropy. Therefore, the diagonal entropy contains the local coherence on each single qubit \cite{local1,local2}, and it should in general scale linearly with the particle number in $\rho _L^{\rm diag}$, i.e., volume effect depending on $L$.
We thus expect that the DE will take the following scaling form on $L$ as
\begin{eqnarray}
\label{Eq:diagentro}
S(\rho _L^{\rm diag})=aL+b\log _2 L+c,
\end{eqnarray}
where the first term is the volume effect with $a$ the DE density in the large $L$ limit, the second term accounts for the logarithmic correction, and the last term is the constant offset.

We also notice that recently the volume law of EE for scramble pure states \cite{Nakagawa2018} and excited eigenstates \cite{Nakagawa2018,excited} of many-body systems are revealed. However, it is well known that the EE in the GSs of 1D models follow area law \cite{arealaw,ZanardiArea,PlenioArea,Hastings,Cirac}. Hence, he volume effect of DE for ground states originates from the quantum coherence (specifically, the local coherence) and is irrelevant to EE.  

In what follows we take an integrable model and one non-integrable model as two examples to compute the GSs, validate Eq. (\ref{Eq:diagentro}) and eventually pinpoint the evidence from the coefficients $\{a,b,c\}$ to identify the quantum phase transitions therein.

\section{Results}\label{S:Result}

\subsection{The XY-model and the ground state diagonal reduced density matrix}
\begin{figure}
%\flushleft
\includegraphics[width=0.38\textwidth]{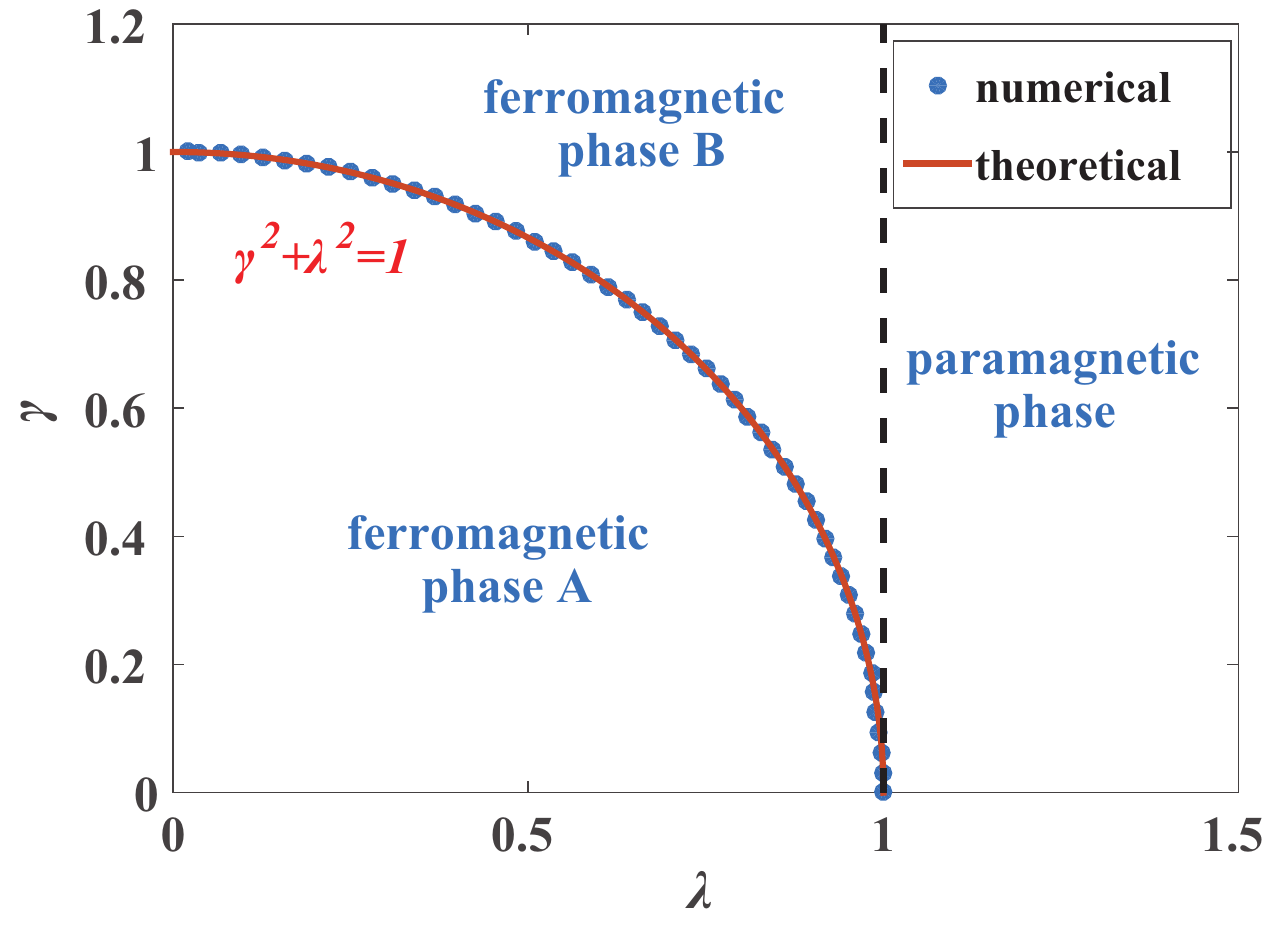}
\caption{Phase diagram of XY-model. The region $0\le \lambda <1$ is ferromagnetic phase,
which is divided into phase A and phase B by relation $\gamma ^2+\lambda ^2=1$. The region $\lambda >1$ is the paramagnetic phase. The critical point for phase transition between ferromagnetic phase and paramagnetic phase is $\lambda =1$, which can be detected by the derivatives of $\{a,b,c\}$, see figure \ref{fig:coefficients}.
The partition between phases A and B can be
determined by condition $c(\lambda )=0$.
The results agree well with the phase diagram of XY-model.}
\label{XYPhaseDiagram}
\end{figure}

We first study the XY-model of spin-$1/2$ chain. The Hamiltonian is written as
$$
H=-\sum_{l=1}^{N}\frac {1}{2}[(1+\gamma)\sigma_{l}^{x}\sigma_{l+1}^{x}+(1-\gamma)\sigma_{l}^{y}\sigma_{l+1}^{y}]+\lambda\sigma_{l}^{z},
\nonumber \\
\label{hamiltonian}
$$
where $\sigma _l^x,\sigma _l^y$ and $\sigma _l^z$ are Pauli matrices and
the subscript $l$ is the site number.
This model describes a variety of XY spin-spin interactions
between nearest neighbours (NN) as well as the effect of an external
magnetic field along the Z direction. It is associated with
several one-dimensional integrable quantum systems.
For $\gamma=0$, the Hamiltonian becomes the XX-model. For $\gamma=\pm1$, it becomes the Ising model with transverse field.
We consider the spin chain in the thermodynamic limit $N\rightarrow \infty$. Periodic boundary condition is assumed.
The phase diagram of XY-model is determined
by both parameters $\gamma$ and $\lambda $, see Refs. \cite{Lieb,McCoy,Katsura}.
The region $\lambda > 1$ is the paramagnetic phase, and the region $\lambda < 1$ is
the ferromagnetic phase which is further divided by the relation $\gamma ^2+\lambda ^2=1$ into two
different phases, ferromagnetic phase A and ferromagnetic phase B, see figure \ref{XYPhaseDiagram}.

We consider a block of $L$ contiguous spins in the chain as party A, which should be translational invariant. The GS of this integrable model can be solved analytically \cite{Lieb,Vidal}, based on which the DE can be calculated with the help of Majorana transformation and Wick theorem \cite{Vidal}.
The details are as following. For convenience, we label the spins in party A as $l = 1,... ,L$.
The corresponding reduced density matrix, $\rho _L$, for the GS of the Hamiltonian can be
expanded in terms of Pauli matrices and the identity $\sigma _l^0$ as
\begin{eqnarray}
\rho_{L}=2^{-L}\sum_{\mu_{1},\dots,\mu_{L}=0,x,y,z}\rho_{\mu_{1}\dots\mu_{L}}\sigma_{1}^{\mu_{1}}\dots\sigma_{L}^{\mu_{L}},\label{rho_z}
\end{eqnarray}
where $\rho_{\mu_{1}\dots\mu_{L}}=\langle\sigma_{1}^{\mu_{1}}\dots\sigma_{L}^{\mu_{L}}\rangle $ is the GS expectation value.
To calculate DE only the diagonal elements are necessary, and we just need to consider the terms, $\mu _l=\{0,z\}$.
Hereafter, we focus on the diagonal matrix $\rho_{L}^{\rm diag}$ with the
corresponding coefficients $\rho_{\mu_{1}\dots\mu_{L}}^{\rm diag}$, which means that
condition $\mu _l=\{0,z\}$ is taken. Explicitly, we have
\begin{eqnarray}
\rho_{\mu_{1}\dots\mu_{L}}^{\rm diag}=\langle\sigma_{1}^{\mu_{1}}\dots\sigma_{L}^{\mu_{L}}\rangle ,~~~\mu _l=\{0,z\}.
\label{correlation_z}
\end{eqnarray}
We then introduce the Majorana operators defined as \cite{Vidal},
\begin{eqnarray}
c_{2l}=\bigg(\prod_{m=0}^{l-1}\sigma_{m}^{z}\bigg)\sigma_{l}^{x},\ \ \   c_{2l+1}=\bigg(\prod_{m=0}^{l-1}\sigma_{m}^{z}\bigg)\sigma_{l}^{y}.
\end{eqnarray}
Then the operator $\sigma ^z_l$ can be written by Majorana operators as
\begin{eqnarray}
\sigma_{l}^{z}=-i\sigma_{l}^{x}\sigma_{l}^{y}=-ic_{2l}c_{2l+1}.
\end{eqnarray}
Substituting this representation into Eq. (\ref{correlation_z}), the coefficients $\rho_{\mu_{1}\dots\mu_{L}}^{\rm diag}$ can be calculated.
The procedure is to calculate the GS expectation values.
For example, for the case $\mu_{m}, \mu_{n}=z$ and $\mu_{l\ne m,n}=0$, the coefficient is rewritten as
\begin{eqnarray}\label{coeff}
\rho_{0\dots 0z0\dots  0z0\dots 0}&=&\langle\sigma_{1}^{0}\dots\sigma_{m}^{z}\dots\sigma_{n}^{z}\dots\sigma_{L}^{0}\rangle\nonumber \\
&=&\langle (-i)^{2}c_{2m}c_{2m+1}c_{2n}c_{2n+1}\rangle \nonumber \\
&=&(-i)^{2}(\langle c_{2m}c_{2m+1}\rangle\langle c_{2n}c_{2n+1}\rangle
\nonumber \\
&&-\langle c_{2m}c_{2n}\rangle\langle c_{2m+1}c_{2n+1}\rangle\nonumber \\
&&+\langle c_{2m}c_{2n+1}\rangle\langle c_{2m+1}c_{2n}\rangle).
\end{eqnarray}
In the last equation, Wick theorem is applied.
Each expectation value in Eq. (\ref{coeff}) can be obtained by using the correlation matrix which is given as
follows, see \cite{Lieb,Vidal},
\begin{eqnarray}
\langle c_{m}c_{n}\rangle=\delta_{mn}+i(\Gamma_{L})_{mn}
\end{eqnarray}
where
\begin{eqnarray}
\Gamma_{L}&=&\left[ \begin{array}{cccc}
\Pi_{0} &\Pi_{1} & \ldots & \Pi_{L-1}\\
\Pi_{-1} & \Pi_{0} &  &\vdots \\
\vdots &   & \ddots & \vdots \\
\Pi_{1-L} & \dots & \dots & \Pi_{0}
\end{array} \right],
\nonumber \\
\Pi_{l}&=&\left[ \begin{array}{cc}
0 & g_{l}\\
-g_{-l} & 0
\end{array} \right].
\end{eqnarray}
For an infinite spin chain, $g_{l}$ takes the form,
\begin{align}
g_{l}=\frac{1}{2\pi}\int_{0}^{2\pi}\ud\phi e^{-il\phi}\frac{\cos\phi-\lambda-i\gamma\sin\phi}{|\cos\phi-\lambda-i\gamma\sin\phi|}.
\end{align}
We notice that the diagonal elements of the matrix $\Pi_{l}$ are all zero,
so the expectation values of two odd or even operators, for example, $\langle c_{2m}c_{2n}\rangle$ or $\langle c_{2m+1}c_{2n+1}\rangle$, are vanishing.
Now, Eq. (\ref{coeff}) can be written as
\begin{align}
\rho_{0\dots z\dots z\dots 0}=g_{0}^{2}-g_{n-m}g_{m-n}.
\end{align}
Summarizing the above calculations, we can derive the diagonal reduced density matrix as
\begin{widetext}
	\begin{align}
	\rho_{L}^{\rm diag}&=2^{-L}[(\sigma_{1}^{0}\dots\sigma_{L}^{0})+\sum_{n=1}^{L}g_{0}(\sigma_{1}^{0}\dots\sigma_{n}^{z}\dots\sigma_{L}^{0})
	+\frac{1}{2!}\sum_{m\ne n}^{L}(g_{0}^{2}-g_{n-m}g_{m-n})(\sigma_{1}^{0}\dots\sigma_{m}^{z}\dots\sigma_{n}^{z}\dots\sigma_{L}^{0})\nonumber\\
	&+\frac{1}{3!}\sum_{l\ne m\ne n}^{L}(g_{0}^{3}-g_{0}(g_{m-l}g_{l-m}+g_{n-m}g_{m-n}+g_{l-n}g_{n-l})+g_{l-m}g_{m-n}g_{n-l}+g_{m-l}g_{n-m}g_{l-n})\nonumber\\
	&(\sigma_{1}^{0}\dots\sigma_{l}^{z}\dots\sigma_{m}^{z}\dots\sigma_{n}^{z}\dots\sigma_{L}^{0})+\dots +\frac{1}{L!}g(\sigma_{1}^{z} \dots\sigma_{L}^{z})],
	\label{diagmatrix}
	\end{align}
\end{widetext}
where the intermediate terms and the explicit form of $g$ in the last term are omitted. However,
all of them can be obtained explicitly based on the rules presented above and will be computed in our numerical calculations.

\begin{figure}
\flushleft
\includegraphics[width=0.45\textwidth,height=0.3\textwidth]{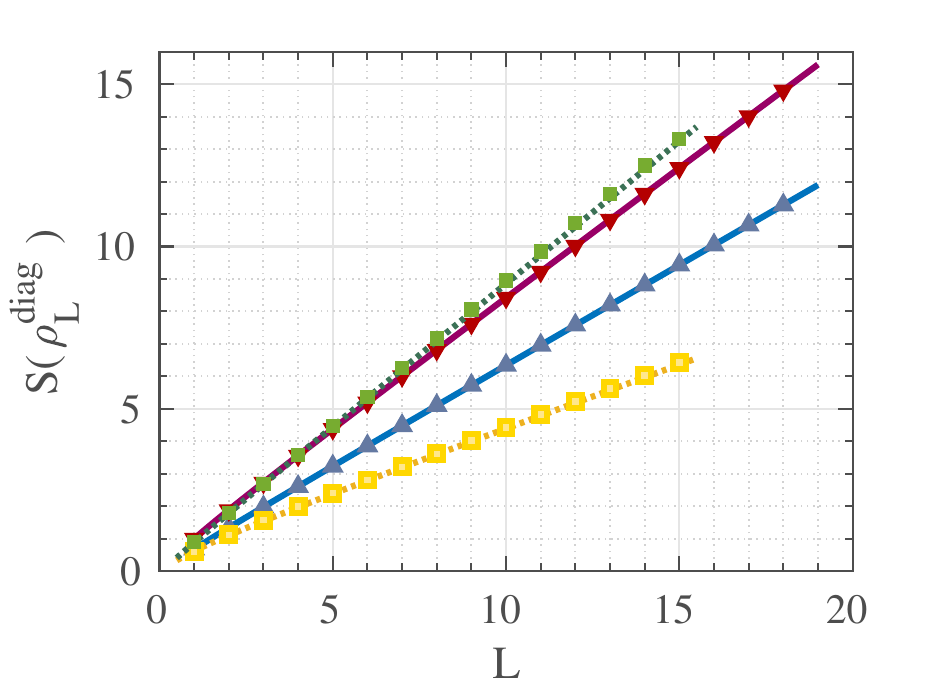}
\caption{DE for a subsystem of contiguous $L$ spins in
the XY-model (red triangles for $\{\gamma=0,\lambda=0 \}$ and blue triangles for $\{\gamma=1,\lambda=1 \}$) and Ising model with NNN interactions (green squares for $J_2=0.1$ and yellow squares for $J_2=0.3$). The solid curves (for XY model) and dotted curves (for NNN interacting Ising model) are correspondingly fitted according to equation (\ref{Eq:diagentro}).}\label{fit18}
\end{figure}

\begin{figure}
\includegraphics[width=0.52\textwidth]{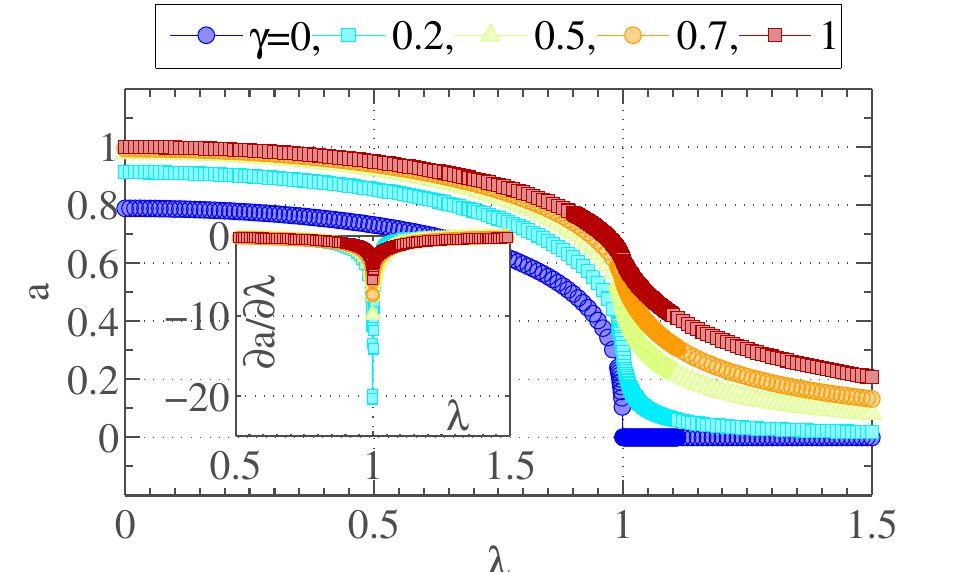}
\includegraphics[width=0.52\textwidth]{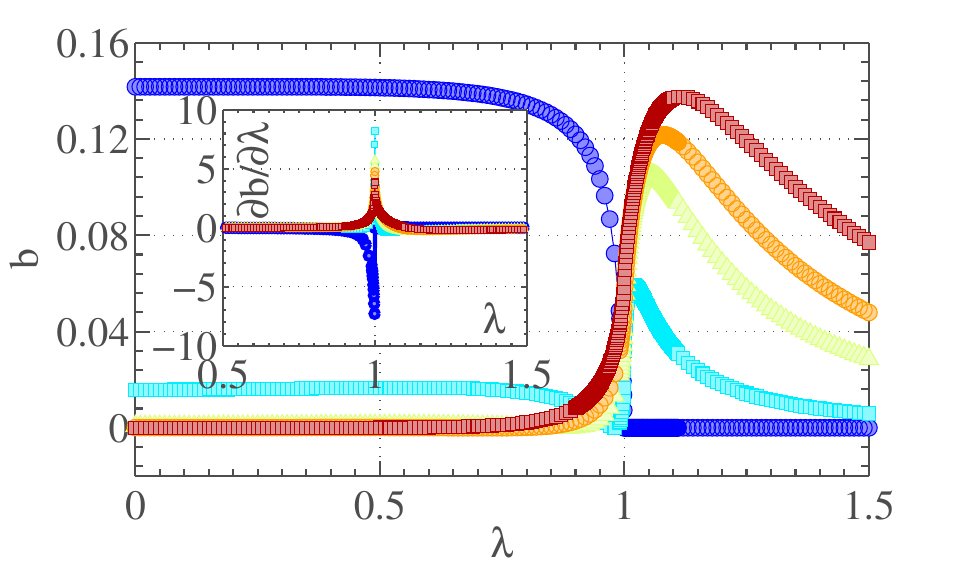}
\includegraphics[width=0.52\textwidth]{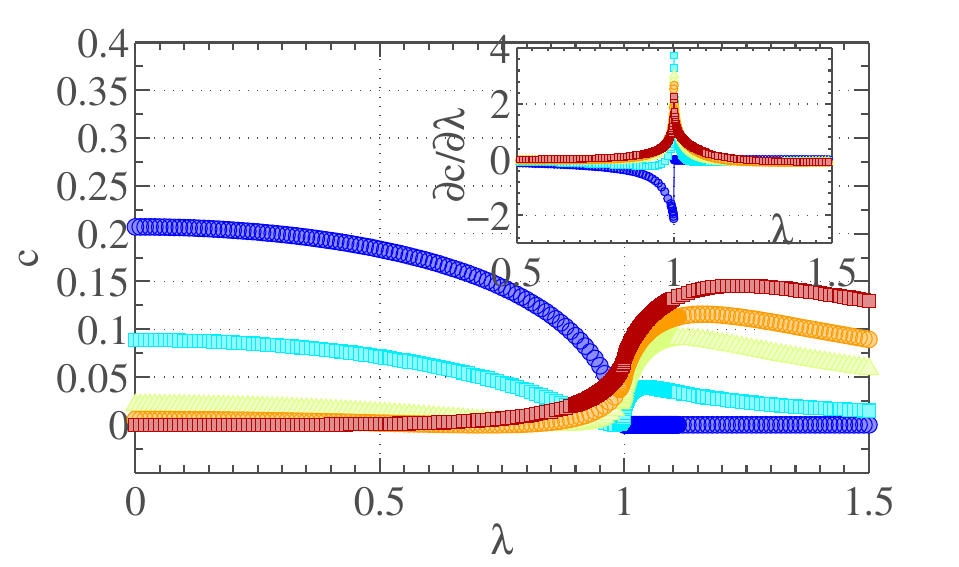}
\caption{ Coefficients in DE of XY-model and the derivatives.
(a) Left is the coefficient $a(\lambda )$, which is the factor in the first term of DE, right
is its derivative. (b) Left is the coefficient $b(\lambda )$ of logarithmic term of DE,
right is its derivative. (c) Left is the constant term $c(\lambda )$ of DE, right is its
derivative. }\label{fig:coefficients}
\end{figure}

For each given GS in the phase diagram we calculate its DE of various $L$ and then fit the data with Eq. (\ref{Eq:diagentro}). This procedure is done when fixing $\gamma$ and sweeping the phase diagram along the direction of increasing $\lambda$ . It is then repeated for $\gamma =0,0.2,0.5,0.7,1.0$.
We find that the volume law of DE is always satisfied, see some examples with $\gamma =0$, $\lambda =0$ (red curve) and $\gamma =1$, $\lambda =1$ (blue curve) in figure \ref{fit18}. Moreover,  $\{a,b,c\}$ can be obtained with high precision when sweeping the phase diagram, see figure \ref{fig:coefficients}. Thus we have shown that the leading term of the DE satisfies the volume law for the whole region of the phase diagram and equation (\ref{Eq:diagentro}) demonstrates precisely the behaviors of DE for GS of XY-model.
We remark that the constant term for Shannon entropy was studied in Refs.\cite{Stephan1,Stephan2,Stephan3} in a different regime of the phase diagram.

%%%%%%%%%%%%%%%%%%%%%%%%%%%
%%%%%%%%%%%%%%%%%%%%%%%%%%%

%%%%%%%%%%%%%%%%%%%%%%%%%%%%

In order to observe explicitly the quantum phase transition at critical point $\lambda =1$, we
calculate the derivatives of those parameters, $\frac {\partial a}{\partial \lambda },
\frac {\partial b}{\partial \lambda }$ and $\frac {\partial c}{\partial \lambda }$.
Remarkably, the quantum phase transitions at $\lambda =1$ can be explicitly identified by all three parameters.
Their derivatives tend to infinity at the critical point, as shown in the insets of figure \ref{fig:coefficients}.
In this sense, DE indeed contains intrinsically the signatures of quantum phase transition between
ferromagnetic phase and paramagnetic phase.

In addition, in the ferromagnetic phase $\lambda <1$,
except for a small region near $\gamma =0$, we notice that
parameter $b$ always tends to be zero,
as shown in panel (b) of figure \ref{fig:coefficients}.
It means that DE may obey the volume law with vanishing logarithm term for most part of the ferromagnetic phase. The nonzero $b$ in the small region near $\gamma=0$ is due to the closing of energy gap. From the inequality $S(\rho_{2L}||\rho_L\otimes\rho_L)\ge S(\mathcal{N}(\rho_{2L})||\mathcal{N}(\rho_L\otimes\rho_L))$ \cite{NielsenChuang}, where $S(\cdot||\cdot)$ is the relative entropy and $\mathcal{N}(\cdot)$ denotes completely positive trace preserving (CPTP) operations, one yields that $2S(\rho_L)-S(\rho_{2L})\ge 2S(\rho_L^{\rm diag})-S(\rho_{2L}^{\rm diag})$. Therefore, if the $1$D system is gapped, where the area law is satisfied \cite{PlenioArea}, the upper bound of $2S(\rho_L^{\rm diag})-S(\rho_{2L}^{\rm diag})$ is a constant $\forall L$ and one can easily derive that $b$ has to be zero, whereas, if the $1$D system is critical like the region $\gamma=0$ in the ferromagnetic phase, the area law no longer holds and as a consequence the possibility of nonvanishing $b$ exists.

Figure \ref{XYPhaseDiagram} shows that the partition between ferromagnetic phase A and ferromagnetic phase B can be located by the condition $c=0$. This result can also be observed
in figure \ref{fig:coefficients} (c). In figure \ref{XYPhaseDiagram},
we present more numerical data for both parameters $\gamma$ and $\lambda $. One can find that the
dots determined by $c=0$ from the results of DE agree well with the known
condition $\gamma ^2+\lambda ^2=1$ (red solid curve), implying that different
phases can be identified by parameters in equation (\ref{Eq:diagentro}). Here the reason is that in case $\gamma ^2+\lambda ^2=1$,
the GSs are product states \cite{XYfactorization1,XYfactorization2}, resulting in both $b=0, c=0$ for DE.
Therefore, numerically by forcing $c(\lambda )=0$ in DE, we can distinguish the two different ferromagnetic phases A and B
to obtain the partition in the phase diagram.

\begin{figure}
	%\flushleft
	\includegraphics[width=0.45\textwidth]{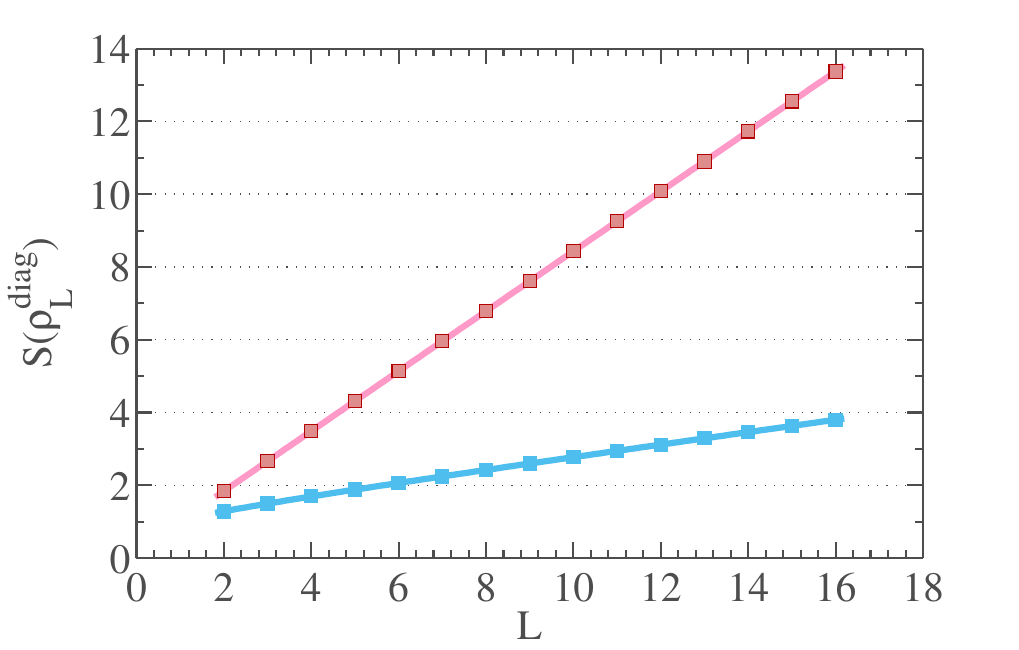}
	\caption{DE of the Ising model GS reduced density matrices with $L$ spins. The dots are the calculated DE values and the solid curves are fitting functions of the form $a\cdot L+b\cdot log(L)+c$. The pink color corresponds the case $\lambda=1.2$ and the blue color corresponds to $\lambda=0.6$. }
	\label{XYvolumelaw_x}
\end{figure}

\begin{figure}
	%\flushleft
	\includegraphics[width=0.45\textwidth]{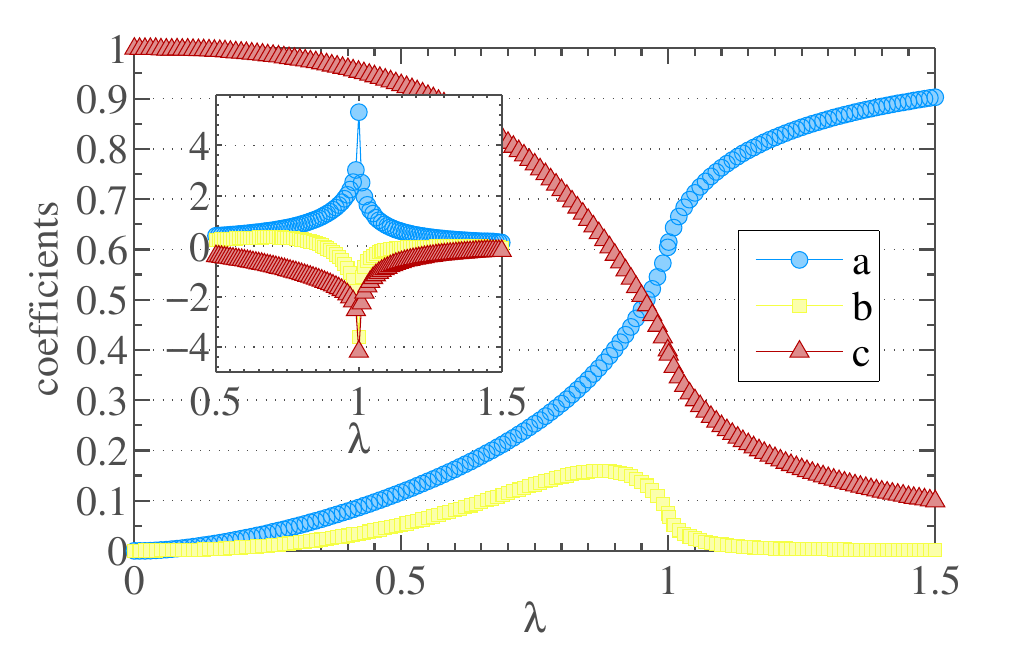}
	\caption{ Coefficients of the fitted DE in the $\sigma_x$ basis when sweeping the phase diagram of Ising model crossing the critical point $\lambda_c=1$. The inset shows the derivatives of the corresponding derivatives with respect to $\lambda$ near the phase transition point.  }
	\label{XYcoefficients_x}
\end{figure}

It is  recognized that the DE is a basis-dependent quantity. Consequently, it is worthwhile to discuss the block scaling law of DE and its capability in detecting quantum phase transitions in other basis. Next, we study the
DE in $\sigma_{x}$ basis, which can be obtained by replacing all $\sigma_z$ components in equations (\ref{rho_z}) and (\ref{correlation_z}).
The expectation values of the $\sigma_x$ components in the thermodynamic limit can be calculated with the cluster decomposition method \cite{Xcomponent}. We show in figure \ref{XYvolumelaw_x} that in the $\sigma_x$ basis the DE obeys the block scaling law (\ref{Eq:diagentro}) as well. Here we consider an Ising model, which corresponds to setting $\gamma=1$ in the XY model. We sweep the phase diagram across the critical point and find the DE volume law is satisfied for all ground states in the $\sigma_x$ basis as well. We then plot the fitted coefficients in figure \ref{XYcoefficients_x}. If one compares it with the result of $\sigma_z$ basis (brown curves in figure \ref{fig:coefficients}), it is quite obvious that although the values of DE as well as the fitted coefficients are different, the singularities in the two bases both occur at the critical point as is evidenced by the derivatives in the inset. This shows that the probe of quantum phase transition does not depend on the choice of basis on which the diagonal entropy is calculated. We can just adopt the computational basis as it is natural for most quantum simulations platform to carry out measurements.

\subsection{Numerical results for non-integrable model}
Next, we consider the quantum Ising model with next nearest neighbour (NNN) interactions as an example to show that the DE volume law also applies to the GSs of non-integrable models. The Hamiltonian reads
$$
 H = -\sum_{i=1}^{L}\sigma_{i}^{z}\sigma_{i+1}^{z}+J_{2}\sum_{i=1}^{L}\sigma_{i}^{z}\sigma_{i+2}^{z}-B_{x}\sum_{i=1}^{L}\sigma_{i}^{x}.
 \label{Eq:HamiltonianIsingNNN}
$$
The first and the second terms are respectively the Ising type interactions between NN and NNN spins, and the last term is the transverse external field. Periodic boundary conditions are assumed, i.e., $\sigma_{L+1}=\sigma_1$ and $\sigma_{L+2}=\sigma_2$. We study the regime of positive $J_2$ yielding a competition between ferromagnetic NN coupling and antiferromagnetic NNN coupling, which ultimately results in the corresponding quantum phase transition in the thermodynamic limit. The whole phase diagram of this model has been studied in Refs. \cite{NNNPhaseDiagramPerturbation, NNNPhaseDiagramDMRG, NNNPhaseDiagramiTEBD} by perturbation theory and numerical simulations with DMRG and iTEBD.
We focus on a line $B_x=0.6$ of the phase diagram.
The quantum phase transition occurs at $\ J_{2c}=0.2$ when scanning the linear path from $J_2=0$ to $ J_2=0.4$.
The frustration makes this model non-integrable and we apply the exact diagonalization method to numerically calculate its ground-state DE with total spin number $N$ up to $16$.
All calculated DE follow the volume law in equation (\ref{Eq:diagentro}). We show two examples with $J_2=0.1$ (green curves) and $J_2=0.3$ (yellow curves) in figure \ref{fit18}.

\begin{figure}
%\flushleft
\includegraphics[width=0.52\textwidth]{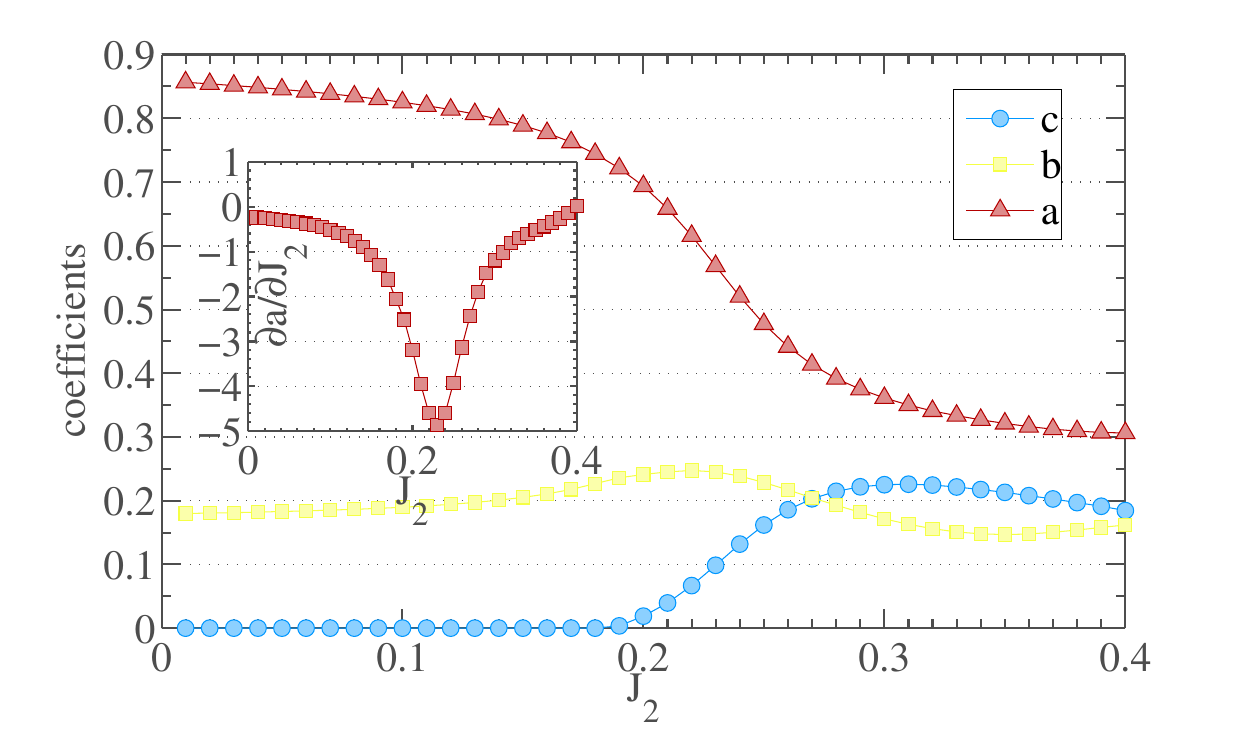}
\caption{ Fitted coefficients of DE for the NNN interacting Ising model with $16$ spins. Inset: the susceptibility of coefficient $a$, i.e., $\partial{a}/\partial{J_2}$.}
\label{fig:NNNIsingN=16}
\end{figure}

\begin{figure}
%\flushleft
\includegraphics[width=0.52\textwidth]{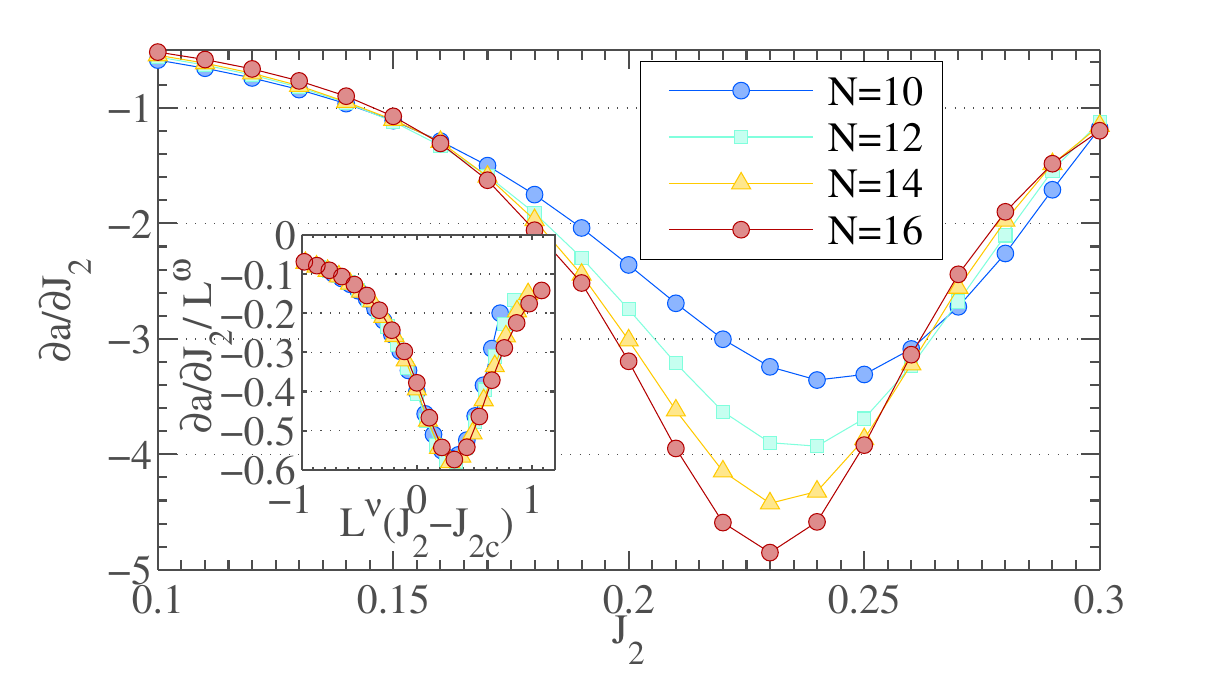}
\caption{Susceptibility of coefficient $a$ for the NNN interacting Ising model with different N. Inset: finite-size scaling in terms of data collapsed according to $\frac{\partial a}{\partial J_{2}} = L^{\omega} f[L^{\nu}(J_{2}-J_{2c})]$. }
\label{fig:NNNIsingdatacollapsed}
\end{figure}

One can see in figure \ref{fig:NNNIsingN=16} that for $N=16$ all three parameters $\{a,b,c\}$ already show obvious indications of phase transition near the critical point: parameter $a$ decreases at its fastest speed, which is evidenced by the derivative in the inset, $b$ arrives at the maximum and $c$ starts to emerge. Notice that all the turning points are slightly away from the phase transition point due to finite size effect. We thus plot $\partial{a}/\partial{J_2}$ for different system sizes in figure \ref{fig:NNNIsingdatacollapsed}, which shows that the peak increases when increasing $N$ and is consistent with a divergence in thermodynamic limit $N\rightarrow\infty$. Moreover, the results in figure \ref{fig:NNNIsingdatacollapsed} indeed demonstrate the trend of moving the minimum point towards $J_{2c}$ with the increase of $N$. The finite-size scaling was carried out in the inset based on the data from $N=10$ up to $N=16$. All curves collapse into one single shape given by the scaling ansatz \cite{CriticalScaling} $\frac{\partial a}{\partial J_{2}} = L^{\omega} f[L^{\nu}(J_{2}-J_{2c})]$ with $\omega = 0.77$, $\nu=0.86$ and $J_{2c}=0.2$.

\section{Discussion}\label{S:Discussion}

In this paper, we find that the universal block scaling law of DE can be represented faithfully as summation of a volume term, a logarithmic term, and a constant term. The involved parameters provide signatures of quantum phase transitions for the studied XY model and the quantum Ising model with next nearest neighbour interactions. It is interesting to generalize the investigation of DE to other quantum phenomena, for instance, topological phase transitions \cite{DEtopology}, dynamical phase transitions \cite{DQPT}, and thermal phase transitions \cite{thermal1,thermal2,thermal3}.

Apart from theoretically being interesting in enhancing our perception of genuine quantumness, DE is also very useful from practical point of view: For experiment of quantum information processing,
a density matrix for an $N-$qubit system can be standardly obtained by state tomography,
however, with exponentially growing effort in the number of qubits for covering all necessary measurement bases \cite{exp4,efftomo}. It is therefore intractable to obtain a density matrix when $N$ is large. Moreover, the off-diagonal entries are quite difficult to measure on many experimental platforms \cite{exp1,exp2}.
In contrast, the diagonal part of the density matrix can be much more easily obtained with merely measurements in the computational basis, which can be translated into the measure of excitation patterns or spin-up-spin-down configurations in atomic platforms. This excellent feature together with the merit of having the potential to characterize a wide range of quantum phenomena makes DE a rather appealing tool of exploring real quantum advantage \cite{Feynman,PreskillQS} on state-of-the-art quantum simulators \cite{Greiner2002,BlochReview,MBLChoi,Browaeys2016Nature,Browaeys2016Science,Bernien2017}.

%%%%%%%%%%%%%%%%%%%%%%%%%%%%%%%%%%%

\begin{acknowledgments}
We thank T. Calarco, J. Rui and A. Rubio Abadal for enlightening
discussions. This work was supported by the
{National Key R \& D Plan of China}
(No. {2016YFA0302104}, No. {2016YFA0300600}),
the {National Natural Science Foundation of China}
(No. {11774406}, {11904018}, {11934018}), the
 {Chinese Academy of Sciences}
 ({XDB28000000}),
 {European Commission}
 funded FET project ``RySQ''
with grant No. {640378},
German Research Foundation
({DFG}) Priority Program
GiRyd, and
{EU} Quantum Flagship project ``PASQuanS''.
\end{acknowledgments}

%\bibliography{reference}
%\bibliographystyle{unsrt}
%

\end{document}